# Interaction between an electric charge and a magnetic dipole of any kind (permanent, para- or dia- magnetic or superconducting)

G. Asti and R. Coïsson
Dept. of Physics and Earth Sciences, University of Parma (Italy)
e-mail giovanni.asti@fis.unipr.it and roberto.coisson@fis.unipr.it

**Abstract**
The interaction between point charge and magnetic dipole is usually considered only for the case of a rigid ferromagnetic dipole (constant-current): here the analysis of force, momentum and energy (including the energy provided by the internal current generator) is generalised to any magnetic dipole behaviour: rigid, paramagnetic, diamagnetic or superconducting (perfectly diamagnetic).

## 1- Introduction

Analysing interactions between point-like electric and/or magnetic systems is a useful exercise for better understanding basic electrodynamic phenomena and their relation with relativistic concepts. In particular even recently there have been controversies about interactions between electric and magnetic particles in connection with concepts like "hidden momentum" [1-13], the force on a magnetic dipole (MD) [14-19] and experiments on the Aharonov-Bohm (AB) and Aharonov-Casher (AC) effects [20-25].
The concept of a point-like electromagnetic system, in particular MDs, is rather delicate as fields go to infinity as the dimensions of the structure goes to zero. A useful model used by Grosse [26] is to extend the spherical shell model used in the first classical models of the electron to a MD: in this case the fields are exactly calculable inside and outside the sphere and the limit to zero radius is easy and unambiguous.
Here we aim to extend the description of interactions between MDs and electric charges to a full range of magnetic behaviours, from a superconducting circuit to a diamagnet and to a rigid (constant current) one and to a paramagnet.
We begin by considering a generic ellipsoid, as this could be useful in order to describe effects with a cigar-like or a disk-like model.

In dealing with the magnet-charge interaction, the exchange of energy has to be studied between the MD, its internal current generator, the charge, and the magnetic and electric fields. The charge is supposed to move with constant velocity on a frictionless rail, under the action of suitable forces. Momentum balance between charge, dipole and field is also considered.
Although the AB (or AC) effect implies a long magnetised filament (or charged filament), the present analysis provides elements of reflection also for these systems, in the limiting case of a constant dipole moment. It seems that the interaction energy in the AB effect, as described in [23], implies a corresponding power to be provided or absorbed by the internal energy source of the magnet, while no work is done on the charge.

We first describe the relation between current, dipole moment, magnetic fields B and H and magnetisation for a model dipole made of surface currents on an ellipsoid, and in particular a sphere.
In general our MD is characterised by a susceptibility χ and a form factor N defining the aspect ratio of the ellipsoid.
Then we analyse forces and work on the charge and on the dipole currents, and the balance of energy of the current generator, the forces needed to keep a constant velocity of the charge, and magnetic and electric field energy, and eventually the momentum balance.
The same problem is briefly considered in the rest system of the charge.

## 2- An ellipsoidal model of a magnetic dipole

The limit to a point-like MD is ambiguous and may be misleading. In order to analyze energies and momenta inside and outside the dipole, a spherical model has been used by Grösse [26]. Proceeding from Grösse's model we try to analyze in detail the fields, energies and momenta involved in the interaction of a MD and an electric charge.
We will use a slightly more general approach, that of an ellipsoid, from which the case of a sphere is immediately derived, and at the same time a needle-like or a disk-like ellipsoid. [27-29]

So we will suppose the MD to be in the form of a small ellipsoid, centered at the origin of an orthogonal reference frame, with semi-axes $a,b$ and $c$, parallel to the axes $x,y$ and $z$ respectively, possibly having axial symmetry around z-axis. In the noticeable case of spherical shape $R=a=b=c$ would be the radius. The surface current density is

$$j = M(\hat{k} \times \hat{n}) \tag{1}$$

where ***n*** is the external unit vector normal to the surface in the specific point and ***k*** is the unity vector of z-axis. The maximum intensity of $j$ is $M$ and occurs at the level of the equatorial plane, coincident with xy plane. Such a body is equivalent to an ellipsoid having a uniform magnetization ***M*** parallel to z-axis so that its magnetic moment turns out to be

$$m = VM \tag{2}$$

where $V = \frac{4}{3}\pi abc$ is the volume of the ellipsoid.

We suppose the magnetization $M$ of the dipole to obey a linear susceptibility equation in dependence of the internal magnetic field $H_i$ (In the geometry that we consider, the external magnetic field is oriented in the same direction of the magnetic moment, so we can use a scalar susceptibility) and to present a spontaneous value $M_r$ (usually called "remnant" or "residual magnetization") at $H_i = 0$, similar to typical ferromagnetic materials.

$$M = \chi H_i + M_r \tag{3}$$

where $\chi$ is the differential susceptibility, while $H_i$ results from the superposition of the external (or "applied") field $H_e = B_e/\mu_0$ and the demagnetizing (or "body") field $H_d = -NM$, $N$ being the so called "demagnetizing factor" relative to z-direction. In these conditions all these internal fields are uniform and parallel to z-axis and linked by the following relations

$$H_i = H_e + H_d = B_e/\mu_0 - NM \quad \text{and} \quad B_i = \mu_0(H_i + M) \ , \tag{4}$$

from which we obtain

$$B_i = B_e + (1-N)I = B_e + B_d \ , \tag{4'}$$

being $I = \mu_0 M$ the "magnetic polarization" and $B_d$ the body induction field.

Now, we can express $I$ as a function of the external induction field $B_e$.

$$I = \frac{\chi}{1+N\chi} B_e + \frac{1}{1+N\chi} I_r = \chi_a B_e + I_a \tag{5}$$

Note that eq. 5 is a general equation, valid for whatever type of magnetic material (dia-, para-, ferro- magnetic as well as superconducting). The coefficient of $B_e$ is the *apparent differential susceptibility* $\chi_a$ while $I_a$ is the *apparent remnant polarization* [30]. Consequently $m$ depends on $B_e$ according to the equation

$$m = \alpha B_e + \frac{1}{1+\chi N} m_r = \alpha B_e + m_a = \Delta m + m_a \ , \tag{6}$$

where $\quad \alpha = \dfrac{V}{\mu_0} \dfrac{\chi}{(1+N\chi)} \tag{6'}$

is the *polarizability* of the magnet, $\quad m_r = V M_r \quad$ and $\quad m_a = V M_a$

Let us remember that $N=1/3$ for a sphere, $N=0$ for a cigar and $N=1$ for a pancake
and $\chi=0$ for a permanent magnet, $\chi>0$ for a paramagnet, $\chi<0$ for a diamagnet, and $\chi=-1$ for a superconductor [see ref. 28 Sec. 5.3.3].

The magnet's field outside a sphere is exactly the dipole field

$$B_d = \frac{\mu_0}{4\pi} m \frac{\hat{e}_\vartheta \sin\vartheta + 2\hat{e}_r \cos\vartheta}{r^3} \ . \tag{7}$$

where $\theta$ and $r$ are spherical coordinates of the field point. Hence the total induction field in any point is the superposition of the external plus the body field $B = B_e + B_d$.

2a- **the case of a superconductor**

In the noteworthy case of a superconducting ellipsoid the induced surface current density is of such intensity to perfectly balance inside it the variation of the external induction field $B_e$. Hence the condition is that

$$\frac{\partial}{\partial B_e} B_i = 0 \ . \tag{8}$$

obtaining

$$M = \frac{-1}{1-N}\frac{B_e}{\mu_0} + \frac{1}{1-N}M_r \tag{9}$$

Note that this equation is in agreement with the general equation 5 for $\chi=-1$, as it should be for a perfect superconductor.

## 3- Interaction of a magnetic dipole with a moving charge

Now let us examine the electromagnetic interaction between the two particles.
We utilize the above described general model, i.e. a surface-current ellipsoidal magnet/circuit ; the aspect ratio depends on the shape-factor $N$ : we have $N=1/3$ for the sphere, and $N$ is close to $1$ for a disk-like ellipsoid and close to zero in the case of a needle-like one.
  The MD $\boldsymbol{m}$ is centered at the origin of an orthogonal reference frame (laboratory frame, $S$) and aligned along the $z$-axis. The charged particle $q$ is moving parallel to $x$-axis at distance $y$ with velocity $\boldsymbol{v}=v\boldsymbol{i}$ in the $xy$ plane and its polar coordinates are $r$ and $\varphi$. It moves on a frictionless rail whose constrictive force compensates the Lorentz force, and a suitable longitudinal force $F_l$ keeps $v$ constant.

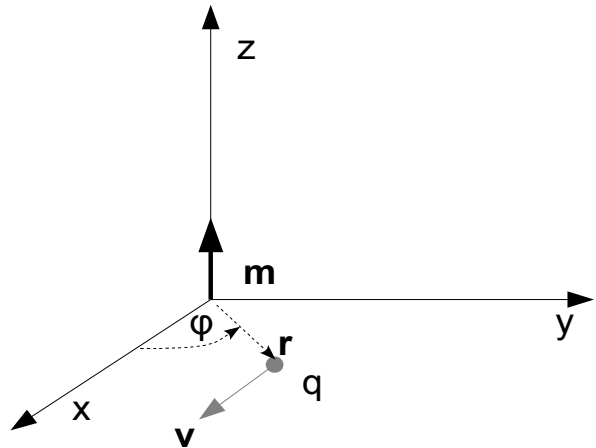

Fig. 1- Scheme of the physical system and Cartesian reference frame

The field yielded by the charge at the dipole position is

$$\boldsymbol{B}_{qm} = \frac{\mu_0(q\boldsymbol{v}\times\boldsymbol{r}^*)}{4\pi r^3} = \frac{\mu_0 qv}{4\pi r^2}\left[\hat{\boldsymbol{i}}\times(-x\hat{\boldsymbol{i}}-y\hat{\boldsymbol{j}})\right] = -\frac{\mu_0 qv}{4\pi r^3}y\hat{\boldsymbol{k}} \tag{10}$$

and then
$$\frac{dm}{dt}=\alpha\frac{\partial B_{qm}}{\partial x}v=\frac{3\mu_0 q\alpha v^2 x y}{4\pi(x^2+y^2)^{5/2}} \tag{11}$$

Here we have utilized $r^*$, the vector from $q$ to $m$, which is the opposite of $r = (x\hat{i} +y\hat{j})r$
And the field produced by the dipole at the charge position is

$$\boldsymbol{B_{mq}}=\nabla\times\boldsymbol{A}=\frac{-\mu_0 m\hat{k}}{4\pi(x^2+y^2)^{3/2}} \tag{12}$$

To the induced magnetic moment $\Delta\boldsymbol{m}$ of the MD corresponds a vector potential

$$\Delta A = \frac{\mu_0(\Delta\boldsymbol{m}\times\boldsymbol{r})}{4\pi r^3} \tag{13}$$

The time variation of $A=A_0+\Delta A$ causes an electric field $E = -\partial A/\partial t$ which in fact accelerate the electric charge $q$.
Then the component of the force along $x$ (i.e. the direction of motion) is

$$F_x=-q\frac{\partial A_x}{\partial t}=\frac{\mu_0 q y}{4\pi(x^2+y^2)^{3/2}}\frac{dm}{dt}=3\alpha x\left(\frac{\mu_0 q v y}{4\pi(x^2+y^2)^2}\right)^2 \tag{14}$$

and along $y$ it is the sum of the Lorentz force and the induction force:

$$F_y=qvB_{mq}-q\frac{\partial A_y}{\partial t}=qvB_{mq}-\frac{\mu_0 q x}{4\pi(x^2+y^2)^{3/2}}\frac{dm}{dt}=\frac{\mu_0 m q v}{4\pi(x^2+y^2)^{3/2}}-3\alpha y\left(\frac{\mu_0 q v x}{4\pi(x^2+y^2)^2}\right)^2 \tag{15}$$

**3a) Energy considerations**

At this point we are interested in analyzing the energy balance in the course of the process. We start considering that the magnetic flux induced in the dipole implies a power of non-electromagnetic nature, $W_m$, to be provided or absorbed by its internal e. m. f. (when available, that is in the case of a para- and ferromagnetic dipole; in the case of a superconducting dipole instead there is no internal extra energy-source that could oppose to the induced flux variation and thus a loss or a gain in the magnetic field energy should result). In order to calculate $W_m$ we cut the ellipsoid in thin slices parallel to $xy$ plane, having thickness $dz$ and crossed on the contour by a current $di$: each one is a an electric circuit of area $\sigma$ and moment $\sigma di$. Then the power feeding the whole pile of circuits is

$$W_m = \int \dot{B}_i \sigma di = \dot{B}_i \int dm = \dot{B}_i m \tag{16}$$

Or else, taking into account that $di=Mdz$ and $\sigma = \pi ab(1-z^2/c^2)$ we can write in more detail

$$W_m = \int_{-c}^{c} dz M \dot{B}_i \pi ab(1-z^2/c^2) = \dot{B}_i \int_{-c}^{c} dz M \pi ab\left(1-\frac{z^2}{c^2}\right) = \dot{B}_i M \pi abc \frac{4}{3} = \dot{B}_i m \tag{17}$$

so that
$$W_m = m\dot{B}_e[1+\chi_a(1-N)] \tag{18}$$

In our case we have $\boldsymbol{B_e}=\boldsymbol{B_{qm}}$. In the case of a permanent magnet $\Delta m=0$ and the internal e. m. f. provides or absorbs a power for keeping the same remnant value of $m$. It is worth noting that if $\boldsymbol{B_{qm}}$ is opposite to $\boldsymbol{m_r}$ (the other side of the path) the power is instead absorbed/provided by the internal generator of the dipole. It is to point out here that, despite the fact that on a mere classical basis the dynamical effects are symmetric on the two paths, the energetic process is instead anti-symmetric.

Let us suppose now, for simplicity, that the charged particle is pushed by external forces in order to maintain a fixed value for its velocity $v$ in the whole journey. Then the power delivered to the traveling charge is

$$W_q = q\mathbf{v} \cdot \mathbf{E} = -q\mathbf{v} \cdot \dot{\mathbf{A}} = -\frac{\mu_0 q\mathbf{v} \cdot (\dot{\mathbf{m}} \times \mathbf{r})}{4\pi r^3} = \frac{\dot{\mathbf{m}} \cdot \mu_0 (q\mathbf{v} \times \mathbf{r})}{4\pi r^3}$$
$$= -\dot{\mathbf{m}} \cdot \mathbf{B}_{qm} = -\alpha \dot{\mathbf{B}}_{qm} \cdot \mathbf{B}_{qm} = -\Delta m \dot{B}_{qm} = -\Delta \dot{m} B_{qm} \qquad (19)$$

So the motion of the charge relative to the MD implies an exchange of energy. Specifically the active force on the charge tends to decelerate it when $|\mathbf{B}_{qm}|$ is increasing – i. e. when $q$ is approaching $\mathbf{m}$ - if $\alpha$ is positive, i. e. when the MD has a para- or ferro-magnetic behaviour. Hence in the case of a superconducting MD $q$ tends to increase its velocity approaching $\mathbf{m}$. As already mentioned eq. 15 tell us that in all cases, differently from the AB effect, the behaviour of the charged particle is perfectly symmetric on opposite paths. Having assumed to keep the charge velocity constant with an external force, it means that this will provide energy in the case of a paramagnet and gain energy in the case of a diamagnet.

But concerning the energetic processes we are especially interested in evaluating the interaction energy between the two particles: this is entirely of magnetic nature as the electric interaction term is vanishing due to symmetry: in fact it is antisymmetric with respect to reflection on the plane containing the z-axis and the charge.

In principle the interaction energy could be determined by integrating the interaction magnetic field term over all space, a calculation which is not an easy task. But it can be deduced utilizing the appropriate energy balance. In fact if we chose a convenient physical system, i.e. the one composed by the charge $q$ and by the electrical currents that form the MD then we can make the following statement: the interaction energy of the magnetic field is increased at the rate $W_m - W_q$ ; it takes the role of internal energy $U$ of the system; here $W_m$ is the power of non-electromagnetic nature received by the system itself, while $W_q$ is the mechanical work per unit time made by the system toward the external world. So the energy balance can be written

$$W_m \delta t = dU + W_q \delta t \quad \text{or} \quad dU/dt = W_m - W_q \qquad (20)$$

Hence we can easily calculate by integration the interaction energy $\Delta U$ of the magnetic field as a function of the particle coordinates $x$ and $y$.

The mechanical energy provided to charge $q$ when coming from infinity to x is

$$L_q = \int_{-\infty}^{t} W_q dt' = -\int_{-\infty}^{t} dt' \dot{m} B_{qm} = -\frac{\alpha}{2} \int_{-\infty}^{t} dt' \frac{\partial B_{qm}^2}{\partial t'} = -\frac{\alpha}{2} B_{qm}^2 = -\frac{1}{2} \Delta m B_{qm} \qquad (21)$$

That is the mechanical energy gained by the charge is equal to one half of the potential magnetic energy of the increment of $m$ in the magnetic field generated by the motion of the charge. Moreover note that if $\alpha=0$ $L_q=0$ (AB effect null ).

The interaction energy we want to calculate is the change in the overall magnetic field energy since the beginning, i.e. when the two particles were infinitely far apart. Then we have

$$\Delta U = \int_{-\infty}^{t}(W_m - W_q)dt' = \int_{-\infty}^{t} m[1 + \chi_a(1-N)]\dot{B}_{qm} dt' + \frac{1}{2}\Delta m B_{qm} = m B_{qm} + \chi_a(1-N)\left[m_a + \frac{1}{2}\Delta m\right] B_{qm} \qquad (22)$$

or $\Delta U = m_a B_{qm}[1+\chi_a(1-N)] + \frac{1}{2}\Delta m B_{qm}[2+\chi_a(1-N)]$ (22')

Note that in the case of a permanent magnet ($\alpha=0; \chi_a=0$) we have $\partial U/\partial t = m \partial B_{qm}/\partial t$ and $\Delta U = m B_{qm}$ while in the case of a superconductor ($\chi_a = (N-1)^{-1}$) we obtain $\partial U/\partial t = \Delta m \partial B_{qm}/\partial t = \alpha B_{qm} \partial B_{qm}/\partial t$ and $\Delta U = \Delta m B_{qm}/2$.

Let us compare this energy with the following expression which is often reported as the "velocity dependent potential" [33]:

$$q\mathbf{v}\cdot\mathbf{A} = \frac{\mu_0 q\mathbf{v}\cdot(\mathbf{m}\times\mathbf{r})}{4\pi r^3} = \mathbf{m}\cdot\frac{\mu_0 q\mathbf{v}\times\mathbf{r}^*}{4\pi r^{*3}} = \mathbf{m}\cdot\mathbf{B}_{qm}$$ (23)

The difference between the two terms gives

$$\Delta U - q\mathbf{v}\cdot\mathbf{A} = \chi_a(1-N)\left[m_a + \frac{1}{2}\Delta m\right]B_{qm}$$ (24)

We observe that this difference is vanishing both when $\chi_a = 0$ and when $N=1$, that is for the permanent MD and for a disk shaped ellipsoid.

**3b) – distribution of magnetic field energy**

The initial magnetic energy is only the creation energy of the MD, that is the energy of its magnetic field in the whole space, which, for the particular case of a spherical MD, is give by [eq.28 in ref. 26]

$$U_0 = \frac{\mu_0 m^2}{4\pi R^3} = \frac{\mu_0 m_a^2}{4\pi R^3} = \frac{\mu_0 m_r^2}{4\pi R^3}$$ (25)

plus the energy of the moving electrical charge.

In a generic instant the magnetic field energy stored in the spherical space inside the MD is

$$U_{B\text{int}} = \frac{4\pi R^3 \cdot B_i^2}{3\cdot 2\mu_0} = \frac{4\pi R^3}{6\mu_0}\left(B_{qm}+\frac{2}{3}\mu_0 M\right)^2 = \frac{4\pi R^3}{6\mu_0}B_{qm}^2 + \frac{4\pi R^3}{3\mu_0}B_{qm}\frac{2}{3}\mu_0 M + \frac{16\pi R^3 \mu_0}{6\cdot 9}M^2$$

$$= \frac{4\pi R^3}{6\mu_0}B_{qm}^2 + \frac{2\cdot 4\pi R^3}{3.3}MB_{qm} + \frac{16\pi^2 R^6 \cdot M^2}{9\cdot \pi R^3 \cdot 6} = \frac{4\pi R^3}{3}\left(\frac{B_{qm}^2}{2\mu_0}\right) + \frac{2}{3}B_{qm}m + \frac{m^2\mu_0}{4\pi R^3}\frac{2}{3}$$ (26)

The last term, which is equal to $(2/3)U_0$, is due to moment itself of the dipole, while the first term is the energy that the external field alone would have in the internal volume of the dipole. The central term is instead part of the interaction energy. Let us consider, as an example, the case of a permanent magnet. We have seen that in this case $\Delta U = m_a B_{qm}$. This is in fact the total interaction energy due to the superposition of the two magnetic fields brought by the two particles. Comparison with eq. (22) tel us that two third of this energy is inside the sphere of the MD while one third is in the external field.

### 3c- Momentum balance

We can see that the force on the magnet $F_{qm}$ is equal and opposite to that on the charge $F_{mq}$, so momentum is conserved.
The force on the magnet is [15,16]:

$$F_{qm} = (\mathbf{m} \cdot \nabla) B_{qm} - \frac{1}{c^2} \left( \frac{d\mathbf{m}}{dt} \times E_{qm} \right) \tag{27}$$

$$F_{qmy} = \frac{-\mu_0 q m v}{4\pi (x^2+y^2)^{3/2}} + \frac{\mu_0 q x}{4\pi (x^2+y^2)^{3/2}} \frac{dm}{dt} \tag{28}$$

$$F_{qmx} = \frac{-\mu_0 q y}{4\pi (x^2+y^2)^{3/2}} \frac{dm}{dt} \tag{29}$$

and these are identical (but with opposite sign) to (14) and (15)
Therefore momentum is conserved as the forces are equal and opposite.

But when the charge has come from infinity to $x=0$, there is now (apart from the initial momentum of the moving charge, which we keep constant), a momentum of the field [31,32]:

$$Q_f = q A = \frac{-\mu_0 m q}{4\pi y^2} \hat{i} \tag{30}$$

but this is compensated by the "hidden momentum" in the magnet:

$$Q_H = \frac{1}{c^2} \mathbf{m} \times E_{qm} = \frac{m q}{4\pi \epsilon_0 c^2 y^2} \hat{i} = -Q_f . \tag{31}$$

### 3d- Angular momentum

Angular momentum (with respect to the position of the dipole) is not conserved, as a moment of the force on the charge is given by the rail forcing it on the line $y$=constant, and the force keeping a constant speed. If we calculate the angular impulse of all forces from $x=-\infty$ to $x=0$ we obtain

$$\Gamma = \int_{-\infty}^{0} F_y x \frac{dx}{v} - \int_{-\infty}^{0} F_x y \frac{dx}{v} = \frac{\mu_0 m q}{4\pi} \int_{-\infty}^{0} \frac{x \, dx}{(x^2+y^2)^{3/2}} + \frac{3\alpha \mu_0^2 q^2 v}{16\pi^2} \left[ y \int_{-\infty}^{0} \frac{x^3 \, dx}{(x^2+y^2)^4} + y^3 \int_{-\infty}^{0} \frac{x \, dx}{(x^2+y^2)^4} \right]$$

$$\rightarrow = \frac{\mu_0 m q}{4\pi y} + \frac{\alpha \mu_0^2 q^2 v}{32\pi^2 y^3} = q A_0 y + q y \Delta A = q A y \tag{32}$$

The first term is due to the Lorentz force and is the total in the case of a rigid magnet ($\Delta A=0$).

## 4- In the reference system of the electric charge.

Looking at the system from a reference where the charge is at rest and the dipole is moving must reach the same results, but the physical description appears different, so it is interesting to mention also this point of view.
The two descriptions might be also relevant to the discussion of the AB and AC effects. In fact, adopting point-like interacting electric and magnetic entities in place of the filamentary bodies considered in the scheme of the AB and AC effects does not change the essential physics (there is no longitudinal force). And the two effects can be considered inherently the same phenomenon.

Hence we will examine the process in another reference system, $S'$, namely that fixed with the electric charge which is set in the point ($x'=a$, $y'=b$, $z'=0$). The two reference frame have axes mutually parallel and the $x'$-axis is sliding on $x$-axis with velocity $v$. Then the MD, as seen from charge $q$, is now traveling at velocity $v^*=-v=-v\hat{i}'$ along the $x'$-axis.
The force that the MD exerts on the charge is the sum of an "electrostatic" force due to the apparent electric dipole due to the motion of the MD and the "induction" force due to the time derivative of the vector potential $A$ [31,32]

The moving magnetic-dipole will generate a transverse electric dipole $p$ given by

$$p = \varepsilon_0\mu_0 v^* \times m = -\varepsilon_0\mu_0 vm(\hat{i}\times\hat{k}) = \varepsilon_0\mu_0 vm\hat{j} \qquad (33)$$

creating on $q$ an electric field and then a force

$$F_{pq} = \frac{\epsilon_0\mu_0 q\, m\, v}{4\pi\epsilon_0 r^3}(3(\hat{j}\cdot\hat{r})\hat{r} - \hat{j}) \qquad (34)$$

Beside this there is also the electric field caused by a time varying vector potential $A$.

$$A = \frac{\mu_0(m\times r)}{4\pi r^3} \quad \text{and} \quad \dot{A} = -\frac{\mu_0(m\times v^*)}{4\pi r^3} + \frac{\mu_0(m\times r)}{4\pi}\frac{d}{dt}r^{-3} + \frac{\mu_0}{4\pi r^3}\frac{dm}{dt}\times r \qquad (35)$$

In the case of a rigid MD, i.e. $dm/dt=0$, the resultant force turns out to be

$$F_{mq} = F_{pq} - q\dot{A} = \frac{\mu_0 q m v}{4\pi r^3}\hat{j} \qquad (36)$$

which is exactly equal to the Lorentz force, as expected.

The magnetic polarizability of $m$ is not evident in this reference frame because charge $q$ does not produce any magnetic field, although we expect to get the same result. This needs a more detailed analysis, which is sketched in the Appendix.

The result is that the MD appears subject to actions equivalent to the presence of the magnetic field $B_{qm}$ that the charge would produce at the position of the MD if it were moving with velocity $v$ (eq. 10).The explanation we give here is a clear solution to the so called "missing torque" paradox reported by Vaidman [15] that gave rise to extensive controversies and discussions in the literature. It refers to the case when a MD is moving in an external electric field.

then $\quad \Delta m = \alpha\, B_{qm} = \frac{\alpha}{c^2} v\times E = \frac{-\alpha q v y}{4\pi\epsilon_0 c^2 r^3}\hat{k} \qquad (37)$

Including then also the third term in the time derivative of $A$, we get an extra force

$$\Delta F_{mq} = \frac{-\mu_0 q}{4\pi r^3}\frac{dm}{dt}\times r \qquad (38)$$

which is identical to eqs. 14-15.

Therefore the momentum balance and energy balance turn out also the same as in sec. 3.

-9-

In order to show that the physical interpretation may be different, let us consider the power exchanged by the internal generator and the field in the case of the rigid MD, which can be calculated as the time derivative of the energy of the equivalent electric dipole in the field $E$ :

$$W = \frac{d}{dt}(\mathbf{p} \cdot \mathbf{E}) = \frac{1}{c^2} \frac{d}{dt}(\mathbf{v}^* \times \mathbf{m} \cdot \mathbf{E}) = \frac{1}{c^2} \frac{d}{dt}(\mathbf{v} \times \mathbf{E} \cdot \mathbf{m}) = m \frac{d}{dt} B_{qm} \tag{39}$$

## 5 - Conclusions

We have described forces, momentum and energy in the quasi-static interaction between a charge and a MD using a spherical or ellipsoidal model, in the case (usually not treated in the literature) that the MD is polarisable, which means it may be a paramagnet, a diamagnet, a ferromagnet with spontaneos moment or a superconducting magnet, or a standard permanent magnet.

In particular a results of this analysis is the energy balance of the system including the energy provided by the internal generator of the dipole. This is often overlooked (except in ref. 23 ) in the analysis of the system composed of a charge and a rigid (i.e. constant dipole moment) MD, and in the case of the Aharonov-Bohm and Aharonov-Casher effects: in fact in that case there is a rate of change of the energy in the field (integrated over all space), which equals the power provided by the internal generator keeping the magnetic moment constant.

## Appendix

How is it possible to have a magnetic susceptibility with the absence of magnetic fields? Well, we should distinguish two cases: a) A MD made by an electrical conductor, say a metal, or a superconductor; b) Dipole made by a magnetic insulator, e. g. a magnetic oxide like magnetite or a ferrite.

In case a) the answer to the question is immediate: the screening charges on the conducting dipole will generate, with their translation at speed $\mathbf{v}^*$, a subsidiary magnetic field, which is exactly the same we observed in the reference frame $S$. The reasons are the following: The induced charges generate an electric field $\mathbf{E}' = -\text{grad}\varphi = -\mathbf{E}$, where $\mathbf{E}$ is the external electric field acting on $\mathbf{m}$. But they generate also a magnetic induction field

$$\mathbf{B}' = \frac{\mu_0}{4\pi} \int \rho_{ind} \mathbf{v}^* \times \frac{\mathbf{r}}{r^3} d\tau = \mathbf{v}^* \times \int \frac{\mu_0 \varepsilon_0}{4\pi \varepsilon_0} \rho_{ind} \frac{\mathbf{r}}{r^3} d\tau = \frac{1}{c^2} \mathbf{v} \times \mathbf{E}' = \mathbf{B} \tag{A1}$$

In case b) we have no magnetic field. However we should remember that the magnetic susceptibility of any magnetic material is a consequence of magnetic anisotropy, of whatever nature, magneto-crystalline, shape, exchange etc..The magnetic anisotropy is an energy density that causes a direct torque on the atomic magnetic moments $\boldsymbol{\mu}$ inside the magnetic material. So the problem is now for us to justify the existence of an equivalent torque on the magnetic moments travelling inside our MD. They are in fact subject to the action of the electric field generated by charge $q$. There are two effects: i) A linear hidden momentum is caused on $\boldsymbol{\mu}_l$,

$$\mathbf{g} = \frac{1}{c^2} \boldsymbol{\mu} \times \mathbf{E} \tag{A2}$$

But there is, associate with $\mathbf{g}$ an angular momentum $\mathbf{G}$ in which the radius vector of $\boldsymbol{\mu}$, $\mathbf{r}_\mu$, is increasing exactly at the speed $\mathbf{v}^*$. As a consequence there is, active on the magnetic moment, a torque ***T*** given by the negative of the time derivative of ***G.***

$$T = -\dot{G} = -v^* \times g = -v^* \times \left(\frac{1}{c^2} \mu \times E\right) \tag{A3}$$

ii) On ***μ*** appears also an electric dipole given by

$$p = \varepsilon_0 \mu_0 v^* \times \mu . \tag{A4}$$

On it the electric field produces a torque

$$L = p \times E = \frac{1}{c^2}(v^* \times \mu) \times E \tag{A5}$$

So the resultant torque acting on ***μ*** is

$$N = T + L = -\dot{G} + L = -v^* \times g + p \times E = -v^* \times \left(\frac{1}{c^2} \mu \times E\right) + \frac{1}{c^2}(v^* \times \mu) \times E$$
$$= \frac{1}{c^2} \mu \times (E \times v^*) = \mu \times B \tag{A6}$$

In the last equation we have made use of the well known cyclic relation regarding the triple vector product. Hence it turns out that ***N*** is equivalent to the effect of the magnetic field as seen from reference system *S*.

## References

[1] term introduced by Shockley and James, but concept used by Poincare' to solve the "4/3" problem in the classical theory of the electron: for this history see ref. [4].

[2] Shockley, W., & James, R. P. (1967). Try simplest cases" discovery of" hidden momentum" forces on" magnetic currents. *Physical Review Letters*, *18*(20), 876.

[3] Coleman, S., & Van Vleck, J. H. (1968). Origin of" Hidden Momentum Forces" on Magnets. *Physical Review*, *171*(5), 1370.

[4] D.J Griffiths, "Resource letter EM1-Electromagnetic Momentum", Am. J. Phys. **80**, 7-18.(2012)

[5] Hnizdo, V. (1992). Conservation of linear and angular momentum and the interaction of a moving charge with a magnetic dipole. *American journal of physics*, *60*(3), 242-246.

[6] Hnizdo, V. (1997). Hidden momentum and the electromagnetic mass of a charge and current carrying body. *American Journal of Physics*, *65*(1), 55-65.

[7] Calkin, MG, (1966), "Linear momentum of quasistatic electromagnetic fields", Amer. J. Phys. 60, 921

[8] V. Hnizdo, "Covariance of the total energy–momentum four-vector of a charge and current carrying macroscopic body", Am. J. Phys. **66**, 414-418 (1998)